\documentclass[12pt,preprint]{aastex}

\begin{document}

\begin{center}
\large\textbf{Electromagnetic instabilities in rotating magnetized viscous objects%
}
\large \\

\smallskip
\smallskip \textsf{A. K. Nekrasov}
\end{center}

Institute of Physics of the Earth, Russian Academy of Sciences, 123995
Moscow, Russia

\noindent e-mail: anatoli.nekrassov@t-online.de

\smallskip 

\noindent \textbf{Abstract. } We study electromagnetic streaming instabilities\ in
thermal viscous regions of rotating astrophysical objects, such as,
magnetized accretion disks, molecular clouds, their cores, and elephant
trunks. The obtained results can also be applied to any regions of
interstellar medium, where different equilibrium velocities between charged
species can arise. \ We consider a weakly ionized multicomponent plasma
consisting of neutrals and magnetized electrons, ions, and dust grains. The
effect of perturbation of collisional frequencies due to density
perturbations of species is taken into account.\ We obtain general
expressions for perturbed velocities of species involving the thermal
pressure and viscosity in the case in which perturbations propagate
perpendicular to the background magnetic field. The dispersion relation is
derived and investigated for axisymmetric perturbations. New compressible
instabilities generated due to different equilibrium velocities of different
charged species are found in the cold and thermal limits when the viscosity
of neutrals can be neglected or is important. The viscosity of magnetized
charged species is negligible for considered perturbations having
wavelengths much larger than the Larmor radius of species. At the same time,
the neutrals are shown to be immobile in electromagnetic perturbations when
their viscosity is sufficiently large.

\noindent Subject headings: accretion, accretion
disks-instabilities-magnetic fields-plasmas-waves

\section{Introduction}

In a series of papers, Nekrasov (2007, 2008 a,b, and 2009 a,b), a general
theory for electromagnetic compressible streaming instabilities in
multicomponent rotating magnetized objects such as accretion disks and
molecular clouds has been developed. For accretion disks, the different
equilibrium velocities of different species (electrons, ions, dust grains,
and neutrals) have been found from the momentum equations taking into
account anisotropic thermal pressure and collisions of charged species with
neutrals (Nekrasov 2008 a,b, 2009 b). For molecular clouds, their cores, and
elephant trunks, the different equilibrium velocities of charged species
have been supposed due to the action of the background magnetic field
(Nekrasov 2009 a). It has been shown that differences of equilibrium
velocities generate compressible streaming instabilities having growth rates
much larger than the rotation frequencies. New fast instabilities found in
these papers have been suggested to be a source of electromagnetic
turbulence in accretion disks and molecular clouds.

Streaming instabilities can also be generated in neutral media. Recently,
the streaming instability originated due to the difference between
equilibrium velocities of small neutral solids and gas has been suggested as
a possible source that could contribute to the planetesimal formation
(Youdin \& Goodman 2005). These authors have numerically treated this
hydrodynamic instability in the Keplerian disk for the interpenetrating
streams coupled via drag forces. The growth rates of instability have been
found to be much smaller than the dynamical timescales. The particle density
perturbations generated by this instability could seed the planetesimal
formation without self-gravity. 

In papers by Nekrasov devoted to electromagnetic streaming instabilities,
the viscosity has not been considered. However, numerical simulations of the
magnetorotational instability show that this effect can influence the
magnitude of the saturated amplitudes of perturbations and, correspondingly,
the turbulent transport of the angular momentum (e.g., Pessah \& Chan 2008;
Masada \& Sano 2008). Interstellar medium, molecular clouds, and accretion
disks around young stars are weakly ionized objects, where collisional
effects play the dominant role. The ratio of the viscosity to the
resistivity (the magnetic Prandtl number) for astrophysical objects takes a
wide range of values. In particular, in accretion disks around compact $X$%
-ray sources and active galactic nuclei, the magnetic Prandtl number varies
by several orders of magnitude across the entire disk (Balbus \& Henry
2008). The viscosity of astrophysical objects is studied mainly in the
framework of the one-fluid magnetohydrodynamics. Therefore, one needs to
consider this effect using a multicomponent fluid approach for studying
streaming instabilities in real multicomponent media.

In the present paper, we investigate electromagnetic streaming
instabilities\ in rotating astrophysical objects, such as, magnetized
accretion disks, molecular clouds, their cores, elephant trunks, and so on,
taking into account effects of collisions, thermal pressure and viscosity.
We consider a weakly ionized multicomponent plasma consisting of electrons,
ions, dust grains, and neutrals. The charged species are supposed to be
magnetized, i.e., their cyclotron frequencies are larger than their orbiting
frequencies and collisional frequencies with neutrals. We take into account
the effect of perturbation of collisional frequencies due to density
perturbations of species, which takes place at different background
velocities of species. We find the general expressions for the perturbed\
velocities of species for perturbations which are perpendicular\ to the
background magnetic field. (The analogous investigation of perturbations
along the magnetic field has been performed in (Nekrasov 2009 c).) These
expressions for any species also contain the perturbed velocity of other
species due to collisions. The dispersion relation is derived and
investigated for axisymmetric perturbations. The cold and thermal regimes
are considered when the pressure force is negligible or dominates the
inertia. The role of the viscosity of neutrals and charged species is
analyzed. The growth rates due to different azimuthal velocities of charged
species are found.

The paper is organized as follows. In Section 2 the basic equations are
given. In Section 3 we shortly discuss the equilibrium state. General
solutions for the perturbed velocities of species for the transverse
perturbations having the radial as well as azimuthal wave number are
obtained in Section 4. These expressions for axisymmetric (radial)
perturbations and magnetized charged species are given in Section 5. In
Section 6 the dispersion relation is derived and its solutions are found.
Discussion of the obtained results and their applicability are given in
Section 7. The main points of the paper are summarized in Section 8.

\section{Basic equations}

In the present paper, we investigate weakly ionized astrophysical objects or
their definite regions, where all the charged species are magnetized, i.e.,
when, in particular, their cyclotron frequencies are larger than their
collisional frequencies with neutrals. Then the momentum equations for
charged species and neutrals including the viscous terms have the form
(Braginskii 1965), 
\begin{eqnarray}
\frac{\partial \mathbf{v}_{j\perp }}{\partial t}+\mathbf{v}_{j}\cdot \mathbf{%
\nabla v}_{j\perp } &=&-\mathbf{\nabla }U-\frac{\mathbf{\nabla }P_{j}}{%
m_{j}n_{j}}+\frac{q_{j}}{m_{j}}\mathbf{E}_{\perp }\mathbf{+}\frac{q_{j}}{%
m_{j}c}\left( \mathbf{v}_{j}\times \mathbf{B}\right) _{\perp }\mathbf{-}\nu
_{jn}\left( \mathbf{v}_{j\perp }-\mathbf{v}_{n\perp }\right)  \\
&&+\mu _{j1}\mathbf{\nabla \nabla \cdot v}_{j}+\mu _{j2}\mathbf{\nabla }^{2}%
\mathbf{v}_{j}\mathbf{\times z+}\mu _{j3}\mathbf{\nabla }^{2}\mathbf{v}%
_{j\perp },  \nonumber
\end{eqnarray}%
\begin{equation}
\frac{\partial v_{jz}}{\partial t}+\mathbf{v}_{j}\cdot \mathbf{\nabla }%
v_{jz}=\frac{q_{j}}{m_{j}}E_{z}\mathbf{+}\frac{q_{j}}{m_{j}c}\left( \mathbf{v%
}_{j}\times \mathbf{B}\right) _{z}\mathbf{-}\nu _{jn}\left(
v_{jz}-v_{nz}\right) +4\mu _{j3}\mathbf{\nabla }^{2}v_{jz},
\end{equation}%
\begin{equation}
\frac{\partial \mathbf{v}_{n}}{\partial t}+\mathbf{v}_{n}\cdot \mathbf{%
\nabla v}_{n}=-\mathbf{\nabla }U-\frac{\mathbf{\nabla }P_{n}}{m_{n}n_{n}}%
-\sum_{j}\nu _{nj}\left( \mathbf{v}_{n}-\mathbf{v}_{j}\right) +\mu
_{n}\left( \mathbf{\nabla }^{2}\mathbf{v}_{n}+\frac{1}{3}\mathbf{\nabla
\nabla \cdot v}_{n}\right) ,
\end{equation}

\noindent where the index $j=e,i,d$ denotes the electrons, ions, and dust
grains, respectively, and the index $n$ denotes the neutrals. In Equations
(1)-(3), $q_{j}$ and $m_{j,n}$ are the charge and mass of species $j$ and
neutrals, $v_{j,n}$ is the hydrodynamic velocity, $n_{j,n}$ is the number
density, $P_{j,n}=n_{j,n}T_{j,n}$ is the thermal pressure, $T_{j,n}$ is the
temperature, $\nu _{jn}$ $=\gamma _{jn}m_{n}n_{n}$ ($\nu _{nj}=\gamma
_{jn}m_{j}n_{j}$) is the collisional frequency of species $j$ (neutrals)
with neutrals (species $j$), where $\gamma _{jn}=<\sigma
v>_{jn}/(m_{j}+m_{n})$ ($<\sigma v>_{jn}$is the rate coefficient for
momentum transfer). The coefficients of the kinematic viscosity are $\mu
_{j1}=c_{j1}\mu _{j}/3$, $\mu _{j2}=c_{j2}\mu _{j}/\omega _{cj}\tau _{jn}$,
and $\mu _{j3}=c_{j3}\mu _{j}/\omega _{cj}^{2}\tau _{jn}^{2}$. Here $\mu
_{j,n}=v_{Tj,n}^{2}/\nu _{jn,nn}$ ($\nu _{nn}$ is the neutral-neutral
collisional frequency), $\omega _{cj}=q_{j}B_{0}/m_{j}c$ is the cyclotron
frequency, $\tau _{jn}=\nu _{jn}^{-1}$, and $v_{Tj,n}=(T_{j,n}/m_{j,n})^{1/2}
$ is the thermal velocity. Numerical coefficients for the electrons and ions
are $c_{e1}=0.73$, $c_{i1}=0.96$, $c_{j2}=0.5$, $c_{e3}=0.51$, and $%
c_{i3}=0.3$ (Braginskii 1965). Further, $U=-GM/R$\ is the gravitational
potential of the central object having mass $M$ (when it presents), $%
R=(r^{2}+z^{2})^{1/2}$, $G$ is the gravitational constant, $E$\ and $B$ are
the electric and magnetic fields, and $c$ is the speed of light in vacuum.
The magnetic field $B$ includes the external magnetic field $B_{0ext}$ of
the central object and/or interstellar medium, the magnetic field $B_{0cur}$
of the stationary current in a steady state, and the perturbed magnetic
field. We use the cylindrical coordinate system $(r,\theta ,z),$ where $r$
is the distance from the symmetry axis $z$, and $\theta $ is the angle in
the azimuthal direction. The background magnetic field is assumed to be
directed along the $z$ axis, $B_{0}=B_{0zext}+B_{0zcur}$, the sign $\perp $
denotes the direction across the $z$ axis. In Equations (1) and (2), the
condition $\omega _{cj}\gg \nu _{jn}$ is satisfied in the viscous terms. For
unmagnetized charged particles of species $j,$ $\omega _{cj}\ll \nu _{jn},$
the viscosity coefficient has the same form as that for neutrals. We adopt
the adiabatic model for the temperature evolution when $P_{j,n}\sim
n_{j,n}^{\gamma _{a}}$, where $\gamma _{a}$ is the adiabatic constant. In
general, we consider the two-dimensional case in which $\nabla =(\partial
/\partial r,\partial /r\partial \theta ,0).$

The other basic equations are the continuity equation, 
\begin{equation}
\frac{\partial n_{j,n}}{\partial t}+\mathbf{\nabla \cdot }n_{j,n}\mathbf{v}%
_{j,n}=0,
\end{equation}

\noindent Faraday`s equation, 
\begin{equation}
\mathbf{\nabla \times E=-}\frac{1}{c}\frac{\partial \mathbf{B}}{\partial t},
\end{equation}%
and Ampere`s law,

\bigskip 

\begin{equation}
\nabla \times B=\frac{4\pi }{c}j,
\end{equation}

\noindent where $j=\sum_{j}q_{j}n_{j}v_{j}.$ We consider the wave processes
with typical time scales much larger than the time the light spends to cover
the wavelength of perturbations. In this case, one can neglect the
displacement current in Equation (6) that results in quasineutrality both
for the electromagnetic and purely electrostatic perturbations.

\section{Equilibrium}

We suppose that in equilibrium the electrons, ions, dust grains, and
neutrals rotate in the azimuthal direction of the astrophysical object
(accretion disk, molecular cloud, its cores, elephant trunk, and so on) with
different, in general, velocities $v_{j,n0}$. These velocities can depend on
the radial coordinate. The stationary dynamics of charged species is
undergone by the action of the background magnetic field and collisions with
neutrals. In their turn, the neutrals also experience a collisional coupling
with charged species influencing their equilibrium velocity. Some specific
cases of equilibrium have been investigated in papers by Nekrasov (2007,
2008 a,b, 2009 b).\ Stationary velocities of charged species in the
background electric field and gravitational field of the central mass in the
absence of collisions have been used in (Nekrasov 2007). The case of weak
collisional coupling of neutrals with light charged species (electrons and
ions) and weak and strong collisional coupling between neutrals and heavy
dust grains has been considered in (Nekrasov 2008 a). Equilibrium where the
neutrals have a strong coupling with light charged species as well as with
heavy dust grains has been investigated in (Nekrasov 2008 b and 2009 b).

In papers cited above, it has been shown that the different charged species
have different stationary velocities. Due to this effect, the electric
currents exist in the equilibrium state, which generate their magnetic
fields.

\section{Linear approximation: General expressions}

In the present paper, we do not treat electromagnetic perturbations
connected with the background pressure gradients. Thus, we exclude the drift
waves from our consideration. We take into account the induced reaction of
neutrals on the perturbed motion of charged species. The neutrals can be
involved in electromagnetic perturbations, if the ionization degree of
medium is sufficiently high. We also include the effect of perturbation on
the collisional frequencies due to density perturbations of charged species
and neutrals. This effect emerges when there are different background
velocities of species. Then the momentum equations (1)-(3) in the linear
approximation take the form, 
\begin{eqnarray}
\frac{\partial \mathbf{v}_{j1\perp }}{\partial t}+\mathbf{v}_{j0}\cdot 
\mathbf{\nabla v}_{j1\perp } &=&-c_{sj}^{2}\frac{\mathbf{\nabla }n_{j1}}{%
n_{j0}}+\frac{q_{j}}{m_{j}}\mathbf{E}_{1\perp }\mathbf{+}\frac{q_{j}}{m_{j}c}%
\left( \mathbf{v}_{j0}\times \mathbf{B}_{1}\right) _{\perp }\mathbf{+}\frac{%
q_{j}}{m_{j}c}\left( \mathbf{v}_{j1}\times \mathbf{B}_{0}\right) _{\perp } \\
&&\mathbf{-}\nu _{jn}^{0}\left( \mathbf{v}_{j1\perp }-\mathbf{v}_{n1\perp
}\right) \mathbf{-}\nu _{jn}^{0}\frac{n_{n1}}{n_{n0}}\left( \mathbf{v}_{j0}-%
\mathbf{v}_{n0}\right) +\mu _{j1}\mathbf{\nabla \nabla \cdot v}_{j1} 
\nonumber \\
&&+\mu _{j2}\mathbf{\nabla }^{2}\mathbf{v}_{j1}\mathbf{\times z+}\mu _{j3}%
\mathbf{\nabla }^{2}\mathbf{v}_{j1\perp },  \nonumber
\end{eqnarray}%
\begin{equation}
\frac{\partial v_{j1z}}{\partial t}+\mathbf{v}_{j0}\cdot \mathbf{\nabla }%
v_{j1z}=\frac{q_{j}}{m_{j}}E_{1z}\mathbf{+}\frac{q_{j}}{m_{j}c}\left( 
\mathbf{v}_{j0}\times \mathbf{B}_{1}\right) _{z}\mathbf{-}\nu
_{jn}^{0}\left( v_{j1z}-v_{n1z}\right) +4\mu _{j3}\mathbf{\nabla }%
^{2}v_{j1z},
\end{equation}%
\begin{eqnarray}
\frac{\partial \mathbf{v}_{n1}}{\partial t}+\mathbf{v}_{n0}\cdot \mathbf{%
\nabla v}_{n1} &=&-c_{sn}^{2}\frac{\mathbf{\nabla }n_{n1}}{n_{n0}}%
-\sum_{j}\nu _{nj}^{0}\left( \mathbf{v}_{n1}-\mathbf{v}_{j1}\right)
-\sum_{j}\nu _{nj}^{0}\frac{n_{j1}}{n_{j0}}\left( \mathbf{v}_{n0}-\mathbf{v}%
_{j0}\right)  \\
&&+\mu _{n}\left( \mathbf{\nabla }^{2}\mathbf{v}_{n1}+\frac{1}{3}\mathbf{%
\nabla \nabla \cdot v}_{n1}\right) ,  \nonumber
\end{eqnarray}

\noindent where $c_{sj,n}=(\gamma _{a}T_{j,n0}/m_{j,n})^{1/2}$ is the sound
velocity ($\gamma _{a}=2(3)$ for the two(one)-dimensional perturbations), $%
\nu _{jn}^{0}=\gamma _{jn}m_{n}n_{n0}$, $\nu _{nj}^{0}=\gamma
_{jn}m_{j}n_{j0}$. The terms proportional to $n_{n1}$ in Equation (7) and $%
n_{j1}$ in Equation (9) describe the effect of perturbation of the
collisional frequencies $\nu _{jn}^{1}=\nu _{jn}^{0}(n_{n1}/n_{n0})$ and $%
\nu _{nj}^{1}=\nu _{nj}^{0}(n_{j1}/n_{j0})$ due to number density
perturbations. The index $1$ denotes quantities of the first order of
magnitude. The neutrals participate in the electromagnetic dynamics only due
to collisional coupling with the charged species. On the left hand-sides of
Equations (7)-(9), we do not take into account the terms of the form $%
v_{j,n1}\cdot \nabla v_{j,n0}$. These terms without neutral dynamics have
been involved in papers by Nekrasov (2007 and 2008 a). The neutral dynamics
has been included in papers (Nekrasov 2008 b and 2009 b). In all specific
cases that were considered the terms mentioned above were negligible. From
general expressions for perturbed velocities given in (Nekrasov (2008 b and
2009 b) it can be seen that these Coriolis terms can be neglected under
sufficient condition $\omega _{j,nz}\gg 2\Omega _{j,n}$, where $\omega
_{j,nz}$\ in the present case are given below and $\Omega _{j,n}$\ are the
rotation frequencies of species in equilibrium. 

The continuity Equation (4) in the linear regime is the following: 
\begin{equation}
\frac{\partial n_{j,n1}}{\partial t}+\mathbf{v}_{j,n0}\cdot \mathbf{\nabla }%
n_{j,n1}+n_{j,n0}\mathbf{\nabla \cdot v}_{j,n1}=0.
\end{equation}

We further apply the Fourier transform to Equations (7)-(10), supposing
perturbations of the form $\exp (ik_{r}r+im\theta -i\omega t).$ Below, we
find expressions for the perturbed velocities of neutrals and charged
species.

\subsection{Perturbed velocity and number density of neutrals}

Using the $z$ component of Equation (9), we find the Fourier amplitude of
the induced longitudinal velocity of neutrals $v_{n1zk}$, 
\begin{equation}
-i\omega _{nz}v_{n1zk}=\sum_{j}\nu _{nj}v_{j1zk},
\end{equation}

\noindent where 
\begin{eqnarray*}
\omega _{nz} &=&\omega _{n}+i\nu _{n}+i\mu _{n}k_{\perp }^{2}, \\
\omega _{n} &=&\omega -\mathbf{k\cdot v}_{n0}, \\
\nu _{n} &=&\sum_{j}\nu _{nj},
\end{eqnarray*}

\noindent $k=\{k,\omega \}=\{k_{r},k_{\theta }=m/r,\omega \}$, $k_{\perp
}^{2}=k_{r}^{2}+k_{\theta }^{2}$. Here and below, we omit, for simplicity,
the index $0$ for $\nu _{jn}^{0}$ and $\nu _{nj}^{0}$.

From Equations (9) and (10), we obtain the induced transverse velocity and
induced perturbed number density of neutrals,\  
\begin{equation}
-i\omega _{nz}\mathbf{v}_{n1\perp k}=\sum_{j}\nu _{nj}\frac{n_{j1k}}{n_{j0}}%
\left[ \mathbf{k}\eta _{n}\frac{\omega _{n}}{\omega _{n\perp }}-\left( 
\mathbf{v}_{n0}-\mathbf{v}_{j0}\right) \right] +\sum_{j}\nu _{nj}\mathbf{v}%
_{j1\perp k},
\end{equation}%
\begin{equation}
-i\omega _{n\perp }\frac{n_{n1k}}{n_{n0}}=\sum_{j}\nu _{nj}\frac{n_{j1k}}{%
n_{j0}},
\end{equation}

\noindent where 
\begin{eqnarray*}
\omega _{n\perp } &=&\omega _{nz}-k_{\perp }^{2}\eta _{n}, \\
\eta _{n} &=&\frac{c_{sn}^{2}}{\omega _{n}}-i\frac{1}{3}\mu _{n}.
\end{eqnarray*}

\noindent We see from Equation (12) that the perturbed velocity of neutrals
is induced by the perturbed number density and velocity of charged species.
Equation (13) shows that the density perturbation of neutrals is defined by
the density perturbations of charged species. When $\omega _{n\perp }\sim
\nu _{nj}$, we obtain $n_{n1k}/n_{n0}\sim n_{j1k}/n_{j0}$. Note that to
obtain the correct expressions (12) and (13) it is necessary to take into
account perturbation on the collisional frequency of neutrals with charged
species.

\ 

\subsection{The longitudinal perturbed velocity of charged species}

Let us now find the longitudinal perturbed velocity of charged species $%
v_{j1z}$. From Equations (5) and (8) in the linear approximation and by
using Equation (11) we obtain, 
\begin{equation}
\omega _{nz}\omega _{jz}v_{j1zk}+\nu _{jn}\sum_{l}\nu _{nl}v_{l1zk}=i\omega
_{nz}F_{j1zk},
\end{equation}

\noindent where 
\begin{eqnarray*}
\omega _{jz} &=&\omega _{j}+i\nu _{jn}+i4\mu _{j3}k_{\perp }^{2}, \\
\omega _{j} &=&\omega -\mathbf{k\cdot v}_{j0}, \\
F_{j1zk} &=&\frac{q_{j}}{m_{j}}\frac{\omega _{j}}{\omega }E_{1zk}.
\end{eqnarray*}

Equation (14) shows that due to collisions of neutrals with charged species
and charged species with neutrals, the charged species experience the
feedback on their perturbations. If neutrals collide with two charged
species only, for example, with electrons and ions, the solutions of
Equation (14) for $e$ and $i$ have the form,

\begin{eqnarray}
D_{z}v_{e1zk} &=&i\alpha _{iz}F_{e1zk}-i\nu _{en}\nu _{ni}F_{i1zk}, \\
D_{z}v_{i1zk} &=&i\alpha _{ez}F_{i1zk}-i\nu _{in}\nu _{ne}F_{e1zk}, 
\nonumber
\end{eqnarray}

\noindent where 
\[
\alpha _{jz}=\omega _{nz}\omega _{jz}+\nu _{jn}\nu _{nj},
\]%
\[
D_{z}=\omega _{nz}\omega _{iz}\omega _{ez}+\omega _{iz}\nu _{en}\nu
_{ne}+\omega _{ez}\nu _{in}\nu _{ni}.
\]

\noindent Substituting $F_{j1zk}$ in Equations (15), we obtain, 
\begin{equation}
D_{z}v_{j1zk}=i\left( a_{jz}-n_{\theta }b_{jz}\right) E_{1zk},
\end{equation}

\noindent where $n_{\theta }=k_{\theta }c/\omega $ and 
\begin{eqnarray*}
a_{ez} &=&\alpha _{iz}\frac{q_{e}}{m_{e}}-\nu _{en}\nu _{ni}\frac{q_{i}}{%
m_{i}}, \\
b_{ez} &=&\alpha _{iz}\frac{q_{e}}{m_{e}}\frac{v_{e0}}{c}-\nu _{en}\nu _{ni}%
\frac{q_{i}}{m_{i}}\frac{v_{i0}}{c}, \\
a_{iz} &=&\alpha _{ez}\frac{q_{i}}{m_{i}}-\nu _{in}\nu _{ne}\frac{q_{e}}{%
m_{e}}, \\
b_{iz} &=&\alpha _{ez}\frac{q_{i}}{m_{i}}\frac{v_{i0}}{c}-\nu _{in}\nu _{ne}%
\frac{q_{e}}{m_{e}}\frac{v_{e0}}{c}.
\end{eqnarray*}

If neutrals collide with ions and dust grains, the index $e$ in Equations
(15) and (16) must be substituted by the index $d$.

\subsection{The transverse perturbed velocity and perturbed number
density of charged species}

We now find the transverse perturbed velocity of charged species $v_{j1k}$.
From Equation (7), we obtain two equations for components $v_{j1rk}$ and $%
v_{j1\theta k},$

\begin{eqnarray}
-i\omega _{j\perp }v_{j1rk} &=&-ik_{r}c_{sj}^{\ast 2}\frac{n_{j1k}}{n_{j0}}%
\mathbf{+}\omega _{cj}^{\ast }v_{j1\theta k}+G_{j1rk}, \\
-i\omega _{j\perp }v_{j1\theta k} &=&-ik_{\theta }c_{sj}^{\ast 2}\frac{%
n_{j1k}}{n_{j0}}\mathbf{-}\omega _{cj}^{\ast }v_{j1rk}+G_{j1\theta k}. 
\nonumber
\end{eqnarray}

\noindent Here, the following notations are introduced:

\begin{eqnarray*}
\omega _{j\perp } &=&\omega _{j}+i\nu _{jn}+i\mu _{j3}k_{\perp }^{2}, \\
\omega _{cj}^{\ast } &=&\omega _{cj}-\mu _{j2}k_{\perp }^{2}, \\
c_{sj}^{\ast 2} &=&c_{sj}^{2}-i\omega _{j}\mu _{j1}, \\
G_{j1r,\theta k} &=&F_{j1r,\theta k}+Q_{j1r,\theta k}, \\
F_{j1rk} &=&\frac{q_{j}}{m_{j}}\left[ E_{1rk}\mathbf{+}\frac{v_{j0}}{c}%
\left( n_{r}E_{1\theta k}-n_{\theta }E_{1rk}\right) \right] , \\
F_{j1\theta k} &=&\frac{q_{j}}{m_{j}}E_{1\theta k}, \\
Q_{j1rk} &=&\nu _{jn}v_{n1rk}, \\
Q_{j1\theta k} &=&\nu _{jn}v_{n1\theta k}\mathbf{-}\nu _{jn}\frac{n_{n1k}}{%
n_{n0}}\left( v_{j0}-v_{n0}\right) .
\end{eqnarray*}

\noindent In the expression for $F_{j1rk}$, we have used Equation (5).
Solutions of Equations (17) have the form, 
\begin{eqnarray}
D_{j\perp }v_{j1rk} &=&i\omega _{j\perp }\left( 1-\delta _{j\theta \theta
}\right) G_{j1rk}+i\left( i\omega _{cj}^{\ast }+\omega _{j\perp }\delta
_{jr\theta }\right) G_{j1\theta k}, \\
D_{j\perp }v_{j1\theta k} &=&i\omega _{j\perp }\left( 1-\delta _{jrr}\right)
G_{j1\theta k}+i\left( -i\omega _{cj}^{\ast }+\omega _{j\perp }\delta
_{jr\theta }\right) G_{j1rk},  \nonumber
\end{eqnarray}

\noindent where 
\[
D_{j\perp }=\omega _{j\perp }^{2}\left( 1-\delta _{j\perp }\right) -\omega
_{cj}^{\ast 2},
\]

\noindent and $\delta _{jlm}=k_{l}k_{m}c_{sj}^{\ast 2}/\omega _{j}\omega
_{j\perp }$, $l,m=r,\theta $, $\delta _{j\perp }=\delta _{jrr}+\delta
_{j\theta \theta }$. The number density perturbation, $n_{j1k}=n_{j0}k\cdot
v_{j1k}/\omega _{j}$, is the following: 
\begin{equation}
\omega _{j}D_{j\perp }\frac{n_{j1k}}{n_{j0}}=\left( ik_{r}\omega _{j\perp
}+k_{\theta }\omega _{cj}^{\ast }\right) G_{j1rk}+\left( ik_{\theta }\omega
_{j\perp }-k_{r}\omega _{cj}^{\ast }\right) G_{j1\theta k}.
\end{equation}

\noindent In the absence of the viscosity, expressions (18) and (19)
coincide with the corresponding expressions in the paper by Nekrasov (2009
b).

\section{Axisymmetric perturbations}

Below, we shall consider axisymmetric perturbations with $k_{r}\neq 0$, $%
k_{\theta }=0$. In this case, Equations (18) and (19) take the form, 
\begin{eqnarray}
D_{j\perp }v_{j1rk} &=&i\omega _{j\perp }\frac{q_{j}}{m_{j}}E_{1rk}+\frac{%
q_{j}}{m_{j}}\left( i\omega _{j\perp }\frac{k_{r}v_{j0}}{\omega }-\omega
_{cj}^{\ast }\right) E_{1\theta k}+i\omega _{j\perp }\nu _{jn}v_{n1rk} \\
&&-\nu _{jn}\omega _{cj}^{\ast }\left[ v_{n1\theta k}\mathbf{-}\frac{n_{n1k}%
}{n_{n0}}\left( v_{j0}-v_{n0}\right) \right] ,  \nonumber \\
D_{j\perp }v_{j1\theta k} &=&\left[ i\omega _{j\perp }\left( 1-\delta
_{jrr}\right) +\omega _{cj}^{\ast }\frac{k_{r}v_{j0}}{\omega }\right] \frac{%
q_{j}}{m_{j}}E_{1\theta k}+\omega _{cj}^{\ast }\frac{q_{j}}{m_{j}}%
E_{1rk}+\omega _{cj}^{\ast }\nu _{jn}v_{n1rk}  \nonumber \\
&&+i\omega _{j\perp }\nu _{jn}\left( 1-\delta _{jrr}\right) \left[
v_{n1\theta k}\mathbf{-}\frac{n_{n1k}}{n_{n0}}\left( v_{j0}-v_{n0}\right) %
\right] .  \nonumber
\end{eqnarray}%
\begin{equation}
\frac{n_{j1k}}{n_{j0}}=\frac{k_{r}v_{j1rk}}{\omega }.
\end{equation}

\noindent From Equation (12), we can find components of the perturbed
velocity of neutrals in the present case, 
\begin{eqnarray}
v_{n1rk} &=&i\frac{1}{\omega _{n\perp }}\sum_{j}\nu _{nj}v_{j1rk}, \\
v_{n1\theta k} &=&-i\frac{1}{\omega _{nz}}\sum_{j}\nu _{nj}\frac{n_{j1k}}{%
n_{j0}}\left( v_{n0}-v_{j0}\right) +i\frac{1}{\omega _{nz}}\sum_{j}\nu
_{nj}v_{j1\theta k}.  \nonumber
\end{eqnarray}%
Substituting these expressions in Equations (20) and using Equations (13)
and (21), we derive equations that contain only perturbed velocities of
charged species, 
\begin{eqnarray}
D_{j\perp }v_{j1rk} &=&i\omega _{j\perp }\frac{q_{j}}{m_{j}}E_{1rk}+\frac{%
q_{j}}{m_{j}}\left( i\omega _{j\perp }\frac{k_{r}v_{j0}}{\omega }-\omega
_{cj}^{\ast }\right) E_{1\theta k}-\frac{\omega _{j\perp }\nu _{jn}}{\omega
_{n\perp }}\sum_{l}\nu _{nl}v_{l1rk} \\
&&-i\frac{\nu _{jn}\omega _{cj}^{\ast }}{\omega _{nz}}\sum_{l}\nu
_{nl}v_{l1\theta k}+i\frac{\nu _{jn}\omega _{cj}^{\ast }}{\omega }%
\sum_{l}\nu _{nl}v_{l1rk}\left[ \frac{k_{r}\left( v_{n0}-v_{l0}\right) }{%
\omega _{nz}}+\frac{k_{r}\left( v_{j0}-v_{n0}\right) }{\omega _{n\perp }}%
\right] ,  \nonumber \\
D_{j\perp }v_{j1\theta k} &=&\left[ i\omega _{j\perp }\left( 1-\delta
_{jrr}\right) +\omega _{cj}^{\ast }\frac{k_{r}v_{j0}}{\omega }\right] \frac{%
q_{j}}{m_{j}}E_{1\theta k}+\omega _{cj}^{\ast }\frac{q_{j}}{m_{j}}E_{1rk} 
\nonumber \\
&&+i\frac{\omega _{cj}^{\ast }\nu _{jn}}{\omega _{n\perp }}\sum_{l}\nu
_{nl}v_{l1rk}-\frac{\omega _{j\perp }\nu _{jn}}{\omega _{nz}}\left( 1-\delta
_{jrr}\right) \sum_{l}\nu _{nl}v_{l1\theta k}  \nonumber \\
&&+\frac{\omega _{j\perp }\nu _{jn}}{\omega }\left( 1-\delta _{jrr}\right)
\sum_{l}\nu _{nl}v_{l1rk}\left[ \frac{k_{r}\left( v_{n0}-v_{l0}\right) }{%
\omega _{nz}}+\frac{k_{r}\left( v_{j0}-v_{n0}\right) }{\omega _{n\perp }}%
\right] .  \nonumber
\end{eqnarray}

We shall find solutions of Equations (23) for the case in which charged
species are magnetized, i.e., when the Lorentz force dominates inertia,
thermal, and drag forces. More specifically, we suppose that 
\begin{equation}
\frac{\omega _{j\perp }}{\omega _{cj}}\left( 1-\delta _{jrr}\right) \ll 
\frac{k_{r}v_{j0}}{\omega }\ll \frac{\omega _{cj}}{\omega _{j\perp }}.
\end{equation}%
It follows from these inequalities that the condition $\omega _{cj}^{2}\gg
\omega _{j\perp }^{2}\left( 1-\delta _{jrr}\right) $ must be satisfied. From
the latter, we obtain that $\omega _{cj}^{2}\gg \omega _{j\perp }^{2}$ and $%
\omega /\omega _{j\perp }\gg k_{r}^{2}\rho _{j}^{2}$, where $\rho
_{j}=c_{sj}/\omega _{cj}$ is the Larmor radius. Thus, the viscosity of
charged species will be negligible for our consideration given below.
Besides, $c_{sj}^{\ast 2}\simeq c_{sj}^{2}$ because we consider that $\omega
\ll \nu _{jn}$. In the case (24), the main perturbed velocities of charged
species (denoted by the upper index $0$) are the electric and magnetic (due
to the Lorentz force) drifts, \  
\begin{eqnarray}
v_{j1rk}^{0} &=&\frac{q_{j}}{m_{j}\omega _{cj}}E_{1\theta k}, \\
v_{j1\theta k}^{0} &=&-\frac{q_{j}}{m_{j}\omega _{cj}}\left( E_{1rk}+\frac{%
k_{r}v_{j0}}{\omega }E_{1\theta k}\right) .  \nonumber
\end{eqnarray}

\noindent Substituting expressions (25) in the collisional terms in
Equations (23), we obtain the following expressions for the transverse
velocities of charged species: 
\begin{eqnarray}
D_{j\perp }v_{j1rk} &=&i\frac{q_{j}}{m_{j}}\sigma _{jrr}E_{1rk}+\frac{q_{j}}{%
m_{j}}\left( -\sigma _{jr\theta 1}+i\sigma _{jr\theta 2}\frac{k_{r}v_{j0}}{%
\omega }+i\sigma _{jr\theta 3}\frac{k_{r}v_{n0}}{\omega }\right) E_{1\theta
k}, \\
D_{j\perp }v_{j1\theta k} &=&\frac{q_{j}}{m_{j}}\sigma _{j\theta r}E_{1rk} 
\nonumber \\
&&+\frac{q_{j}}{m_{j}}\left[ i\sigma _{jr\theta 2}-i\omega _{j\perp }\delta
_{jrr}+\left( \sigma _{jr\theta 1}-\sigma _{j\theta \theta 1}\right) \frac{%
k_{r}v_{j0}}{\omega }+\sigma _{j\theta \theta 2}\frac{k_{r}v_{n0}}{\omega }%
\right] E_{1\theta k},  \nonumber
\end{eqnarray}

\noindent where notations are introduced, 
\begin{eqnarray*}
\sigma _{jrr} &=&\omega _{j\perp }+\frac{\nu _{jn}\nu _{n}}{\omega _{nz}}%
,\sigma _{jr\theta 1}=\omega _{cj}+\frac{\omega _{j\perp }\nu _{jn}\nu _{n}}{%
\omega _{n\perp }\omega _{cj}},\sigma _{jr\theta 2}=\omega _{j\perp }+\frac{%
\nu _{jn}\nu _{n}}{\omega _{n\perp }}, \\
\sigma _{jr\theta 3} &=&\nu _{jn}\nu _{n}\left( \frac{1}{\omega _{nz}}-\frac{%
1}{\omega _{n\perp }}\right) ,\sigma _{j\theta r}=\omega _{cj}+\frac{\omega
_{j\perp }\nu _{jn}\nu _{n}}{\omega _{nz}\omega _{cj}}\left( 1-\delta
_{jrr}\right) , \\
\sigma _{j\theta \theta 1} &=&\frac{\omega _{j\perp }\nu _{jn}\nu _{n}}{%
\omega _{n\perp }\omega _{cj}}\delta _{jrr},\sigma _{j\theta \theta 2}=\frac{%
\omega _{j\perp }\nu _{jn}\nu _{n}}{\omega _{cj}}\left( 1-\delta
_{jrr}\right) \left( \frac{1}{\omega _{nz}}-\frac{1}{\omega _{n\perp }}%
\right) .
\end{eqnarray*}

\noindent Below, we shall derive the dispersion relation and find its
solutions.

\section{Dispersion relation}

From Faraday's and Ampere's laws (5) and (6), we find the following
equations: 
\begin{eqnarray}
j_{1rk} &=&0, \\
n_{r}^{2}E_{1\theta k} &=&\frac{4\pi i}{\omega }j_{1\theta k},  \nonumber \\
n_{r}^{2}E_{1zk} &=&\frac{4\pi i}{\omega }j_{1zk},  \nonumber
\end{eqnarray}

\noindent where $j_{1r,zk}=\sum_{j}q_{j}n_{j0}v_{j1r,zk}$, $j_{1\theta
k}=\sum_{j}q_{j}n_{j0}v_{j1\theta k}+\sum_{j}q_{j}n_{j1k}v_{j0}$, and $%
n_{r}^{2}=k_{r}^{2}c^{2}/\omega ^{2}$. We see that perturbations with
polarizations of the electric field along and across the background magnetic
field are split. We do not consider perturbations with longitudinal
polarization because the current $j_{1zk}$ does not contain the equilibrium
velocities in the axisymmetric case (see Equation (16)). Using expressions
(26), we find the transverse electric currents, 
\begin{eqnarray}
\frac{4\pi i}{\omega }j_{1rk} &=&-\varepsilon _{rr}E_{1rk}-\varepsilon
_{r\theta }E_{1\theta k}, \\
\frac{4\pi i}{\omega }j_{1\theta k} &=&\varepsilon _{\theta
r}E_{1rk}+\varepsilon _{\theta \theta }E_{1\theta k}.  \nonumber
\end{eqnarray}

\noindent Here, 
\begin{eqnarray*}
\varepsilon _{rr} &=&\sum_{j}\frac{\omega _{pj}^{2}}{\omega D_{j\perp }}%
\sigma _{jrr},\varepsilon _{r\theta }=\sum_{j}\frac{\omega _{pj}^{2}}{\omega
D_{j\perp }}\left( i\sigma _{jr\theta 1}+\sigma _{jr\theta 2}\frac{%
k_{r}v_{j0}}{\omega }+\sigma _{jr\theta 3}\frac{k_{r}v_{n0}}{\omega }\right)
, \\
\varepsilon _{\theta r} &=&\sum_{j}\frac{\omega _{pj}^{2}}{\omega D_{j\perp }%
}\left( i\sigma _{j\theta r}-\sigma _{jrr}\frac{k_{r}v_{j0}}{\omega }\right)
, \\
\varepsilon _{\theta \theta } &=&\sum_{j}\frac{\omega _{pj}^{2}}{\omega
D_{j\perp }}\left[ -\sigma _{jr\theta 2}\left( 1+\frac{k_{r}^{2}v_{j0}^{2}}{%
\omega ^{2}}\right) +\omega _{j\perp }\delta _{jrr}+\left( i\sigma _{j\theta
\theta 2}-\sigma _{jr\theta 3}\frac{k_{r}v_{j0}}{\omega }\right) \frac{%
k_{r}v_{n0}}{\omega }\right] .
\end{eqnarray*}%
where $\omega _{pj}=\left( 4\pi n_{j0}q_{j}^{2}/m_{j}\right) ^{1/2}$ is the
plasma frequency.

From Equations (27) and (28), we obtain the dispersion relation, 
\begin{equation}
\varepsilon _{rr}n_{r}^{2}=\varepsilon _{rr}\varepsilon _{\theta \theta
}-\varepsilon _{r\theta }\varepsilon _{\theta r}.
\end{equation}

\noindent Below, we will find solutions of Equation (29) in limiting cases.

\subsection{The case of cold species}

We at first consider the case in which neutrals and charged species are
sufficiently cold, $\omega ^{2}\gg k_{r}^{2}c_{sn}^{2}$ and $\omega \nu
_{jn}\gg k_{r}^{2}c_{sj}^{2}$ (it is obvious that the first condition is the
main one). We also suppose that $\nu _{jn}\gg \nu _{n}\gg \omega $. Under
these conditions the values $\sigma $\ have a simple form, $\sigma
_{jrr}=\sigma _{jr\theta 2}=\omega \nu _{jn}/\nu _{n}$,\ $\sigma _{j\theta
r}=\sigma _{jr\theta 1}=\omega _{cj}+\nu _{jn}^{2}/\omega _{cj}+i\omega \nu
_{jn}^{2}/\omega _{cj}\nu _{n}$, and $\sigma _{jr\theta 3}=\sigma _{j\theta
\theta 2}=0$. When calculating the value $\varepsilon _{rr}\varepsilon
_{\theta \theta }-\varepsilon _{r\theta }\varepsilon _{\theta r}$, we have
carried out symmetrization, used the condition of quasineutrality, $%
\sum_{j}q_{j}n_{j0}=0$, and kept the main terms. Then the dispersion
relation (29) takes the form,

\begin{equation}
\omega ^{2}\sum_{jl}\frac{\omega _{pj}^{2}\nu _{jn}}{\omega _{cj}^{2}}\frac{%
\omega _{pl}^{2}\nu _{ln}}{\omega _{cl}^{2}}=\nu _{n}\sum_{j}\frac{\omega
_{pj}^{2}\nu _{jn}}{\omega _{cj}^{2}}k_{r}^{2}c^{2}-\frac{1}{2}\sum_{jl}%
\frac{\omega _{pj}^{2}\nu _{jn}}{\omega _{cj}^{2}}\frac{\omega _{pl}^{2}\nu
_{ln}}{\omega _{cl}^{2}}k_{r}^{2}\left( v_{j0}-v_{l0}\right) ^{2}.
\end{equation}

\noindent We see that the streaming instability is possible when the
difference of the equilibrium velocities of charged species is sufficiently
large to exceed the threshold defined by the first term on the right-hand
side of Equation (30).

Let us consider, for example, Equation (30) for the electron-ion plasma
(without dust grains). Then this equation will be the following:%
\begin{equation}
\omega ^{2}=k_{r}^{2}c_{A}^{2}-\frac{m_{e}\nu _{en}}{m_{i}\nu _{in}}%
k_{r}^{2}\left( v_{e0}-v_{i0}\right) ^{2},
\end{equation}%
where $c_{A}=\left( B_{0}^{2}/4\pi m_{n}n_{n0}\right) ^{1/2}$ is the Alfv%
\'{e}n velocity. When obtaining this equation, we have used that $m_{i}\nu
_{in}\gg m_{e}\nu _{en}$ (see e.g. Section 7). The first term on the right
hand-side of Equation (31) describes the magnetosonic waves in a weakly
ionized plasma. The second term can generate the streaming instability of
the hydrodynamic kind. This term does not change the order of the dispersion
relation for perturbations under consideration that results in appearance of
the threshold for the absolute instability to develop (see also Nekrasov
2007).

\subsection{The case of thermal species}

In the case of thermal species, we suppose conditions $\mu _{n}k_{r}^{2}\gg
\omega $ and $k_{r}^{2}c_{sj}^{2}\gg \omega \nu _{jn}$ to be satisfied. Then
we have that $k_{r}^{2}c_{sn}^{2}\gg \omega ^{2}$ (the intermediate case for
neutrals, $\mu _{n}k_{r}^{2}\ll \omega \ll k_{r}c_{sn}$, we will not
consider). We also adopt, as above, that $\nu _{jn}\gg \nu _{n}$. However,
the condition $\nu _{n}\gg \omega $ is now not necessary. Then the values $%
\sigma $\ take the form, 
\begin{eqnarray*}
\sigma _{jrr} &=&i\kappa \nu _{jn}\frac{\mu _{n}k_{r}^{2}}{\nu _{n}},\sigma
_{jr\theta 1}=\omega _{cj},\sigma _{jr\theta 2}=i\nu _{jn},\sigma _{jr\theta
3}=-i\kappa \nu _{jn}, \\
\sigma _{j\theta r} &=&\omega _{cj}+\sigma _{j\theta \theta 2},\sigma
_{j\theta \theta 2}=i\kappa \nu _{jn}d_{j},
\end{eqnarray*}%
where $\kappa =\nu _{n}/\left( \nu _{n}+\mu _{n}k_{r}^{2}\right) $\ and $%
d_{j}=k_{r}^{2}c_{sj}^{2}/\omega \omega _{cj}$. Substituting these values in 
$\varepsilon _{rr,\theta }$ and $\varepsilon _{\theta r,\theta }$, we find
expression $\varepsilon _{rr}\varepsilon _{\theta \theta }-\varepsilon
_{r\theta }\varepsilon _{\theta r}$, 
\begin{equation}
\varepsilon _{rr}\varepsilon _{\theta \theta }-\varepsilon _{r\theta
}\varepsilon _{\theta r}=i\kappa \frac{\mu _{n}k^{2}}{\nu _{n}}%
\sum_{jl}\lambda _{j}\lambda _{l}d_{j}+\kappa \frac{\mu _{n}k^{2}}{\nu _{n}}%
\sum_{jl}\lambda _{j}\lambda _{l}\frac{k_{r}^{2}\left( v_{j0}-v_{l0}\right)
^{2}}{2\omega ^{2}},
\end{equation}%
where $\lambda _{j}=-\omega _{pj}^{2}\nu _{jn}/\omega \omega _{cj}^{2}$.
When deriving Equation (32), we have used that $\sum_{j}\omega
_{pj}^{2}\omega _{cj}/\omega D_{j\perp }=-i\sum_{j}\lambda _{j}d_{j}$ due to
condition of quasineutrality. The first term on the right hand-side of this
equation is the main one according to conditions at hand among other terms
that are independent of background velocities. We see from Equation (32)
that the viscosity of neutrals plays an important role in the case under
consideration. 

The dispersion relation (29) takes the form,%
\begin{equation}
\omega \left[ 2\sum_{j}\frac{\omega _{pj}^{2}\nu _{jn}}{\omega _{cj}^{2}}%
c^{2}+\sum_{jl}\frac{\omega _{pj}^{2}}{\omega _{cj}^{2}}\frac{\omega
_{pl}^{2}}{\omega _{cl}^{2}}\left( \nu _{ln}c_{sj}^{2}+\nu
_{jn}c_{sl}^{2}\right) \right] =i\sum_{jl}\frac{\omega _{pj}^{2}\nu _{jn}}{%
\omega _{cj}^{2}}\frac{\omega _{pl}^{2}\nu _{ln}}{\omega _{cl}^{2}}\left(
v_{j0}-v_{l0}\right) ^{2}
\end{equation}

\noindent Thus, the streaming instability exists for the sufficiently short
wavelengths of perturbations when the viscosity of neutrals and thermal
pressure of species play a role.

We now consider Equation (33) for the electron-ion plasma (as above). In
this case, we obtain,%
\begin{equation}
\omega =i\frac{m_{e}}{m_{i}}\nu _{en}\frac{\left( v_{e0}-v_{i0}\right) ^{2}}{%
\left( c_{Ai}^{2}+c_{s}^{2}\right) },
\end{equation}%
where $c_{Ai}=\left( B_{0}^{2}/4\pi m_{i}n_{i0}\right) ^{1/2}$ is the ion
Alfv\'{e}n velocity and $c_{s}=\left[ \gamma _{a}\left( T_{e}+T_{i}\right)
/m_{i}\right] ^{1/2}$ is the ion sound velocity. We see from Equation (34)
that the growth rate $\gamma =$Im $\omega $ can be sufficiently large in
comparison to the rotation frequency in media where $c_{Ai}^{2}+c_{s}^{2}$
is not too large as compared to $\left( v_{e0}-v_{i0}\right) ^{2}$.

The instability described by Equation (33) belongs to the class of
dissipative instabilities because the dispersion relation contains the
corresponding imaginary term which is proportional to the collisional
frequency. Without additional damping mechanisms, the threshold of this
dissipative-streaming instability is absent. 

\section{Discussion}

Let us discuss some consequences that follow from the results obtained
above. From Equation (13), we see that the relative number density
perturbation of neutrals is of the order of that for charged species, if $%
\omega _{n\perp }\sim \nu _{n}$. However, in the case $k^{2}c_{sn}^{2}\gg
\omega \nu _{n}$, a neutral fluid can be considered as incompressible. In
the latter case, the perturbed radial velocity of neutrals in the
axisymmetric perturbations is also negligible (see Equation (22)). At the
same time, the perturbed azimuthal velocity of neutrals is of the order of
that for charged species for small viscosity of neutrals, $\nu _{n}\gg \mu
_{n}k_{r}^{2}$ $(\nu _{n}\gg \omega )$, and is negligible for large
viscosity of neutrals, $\nu _{n}\ll \mu _{n}k_{r}^{2}$ (see Equation (22)).
Thus, the neutrals are immobile in perturbations when their viscosity is
large.

As a specific example, we consider the case of the dense regions of
molecular cloud cores where the number density of neutrals is $n_{n}\sim
10^{5}$ cm$^{-3}$ and $n_{i}/n_{n}\sim 10^{-7}$ (Caselli et al. 1998; Ruffle
et al. 1998). For simplicity, we omit the index $0$ by $n_{j0}$. The ratio
of mass densities of dust grains, $\rho _{d}=m_{d}n_{d}$, and neutrals, $%
\rho _{n}=m_{n}n_{n}$, in the interstellar medium is of the order of $10^{-2}
$ (e.g. Abergel et al. 2005). From these relations, we obtain that $n_{d}\ll
n_{i}$ at $m_{d}\gg 10^{5}m_{n}$. For usually adopted $m_{n}=2.33m_{p}$,
where $m_{p}$ is the proton mass, and for $\sigma _{d}=1$ g cm$^{-3}$, where 
$\sigma _{d}$ is the density of grain material, we obtain that $m_{d}\gg
10^{5}m_{n}=3.9\times 10^{-19}$ g is satisfied at $r_{d}$ $\gg 4.53\times
10^{-3}$ $\mu $m, where $r_{d}$ is the grain radius. The typical values of
grain radius are $r_{d}$ $\sim 0.01-1$ $\mu $m (Mendis \& Rosenberg 1994;
Wardle \& Ng 1999). However, in spite of that $n_{d}\ll n_{i}$ the mass
density of dust grains $\rho _{d}\sim 7.77\times 10^{3}\rho _{i}$ for $%
m_{i}=30m_{p}$.

Further, we take the magnetic field $B_{0}=150$ $\mu $G and the radius and
charge of dust grains $r_{d}=0.01$ $\mu m$ and $q_{d}=q_{e}$ (we consider
one typical type of dust grains). In this case, $m_{d}=4.19\times 10^{-18}$
g and $n_{d}=0.93\times 10^{-3}$ cm$^{-3}$. Then we obtain the following
values for the plasma and cyclotron frequencies of charged species: $\omega
_{pe}=5.64\times 10^{3}$ s$^{-1}$, $\omega _{ce}(>0)=2.64\times 10^{3}$ s$%
^{-1}$, $\omega _{pi}=24$ s$^{-1}$, $\omega _{ci}=4.78\times 10^{-2}$ s$^{-1}
$ ($q_{i}=-q_{e}$), $\omega _{pd}=2.53\times 10^{-2}$ s$^{-1}$, and $\omega
_{cd}(>0)=5.75\times 10^{-7}$ s$^{-1}$. The rate coefficients for momentum
transfer by elastic scattering of electrons and ions with neutrals are $%
<\sigma \nu >_{en}=4.5\times 10^{-9}(T_{e}/30$ K$)^{1/2}$ cm$^{3}$ s$^{-1}$
and $<\sigma \nu >_{in}=1.9\times 10^{-9}$ cm$^{3}$ s$^{-1}$ (Draine et al.
1983). We take $T_{e}=300$ K. Then we obtain $\nu _{en}$ $=1.42\times 10^{-3}
$ s$^{-1}$ and $\nu _{in}$ $=1.37\times 10^{-5}$ s$^{-1}$. The collisional
frequency of dust grains with neutrals has the form $\nu _{dn}\simeq 6.7\rho
_{n}r_{d}^{2}v_{Tn}/m_{d}$ ($T_{n}\sim T_{e}$) (e.g., Wardle \& Ng 1999).
Using parameters given above, we obtain $\nu _{dn}=6.24\times 10^{-8}$ s$%
^{-1}$. The collisional frequency of neutrals with charged species is $\nu
_{n}\simeq \nu _{nd}=6.24\times 10^{-10}$ s$^{-1}$.

The values $\omega _{pj}^{2}\nu _{jn}/\omega _{cj}^{2}$ which are contained
in Equations (30) and (33) are under conditions at hand the following: $%
\omega _{pe}^{2}\nu _{en}/\omega _{ce}^{2}=6.4\times 10^{-3}$ s$^{-1}$, $%
\omega _{pi}^{2}\nu _{in}/\omega _{ci}^{2}=3.44$ s$^{-1}$, $\omega
_{pd}^{2}\nu _{dn}/\omega _{cd}^{2}=120.8$ s$^{-1}$. Thus, we can write
Equation (30) in the form, 
\[
\omega ^{2}=\frac{\nu _{nd}}{\nu _{dn}}\left[ 1-\frac{\nu _{in}}{\nu _{nd}}%
\frac{\left( v_{i0}-v_{d0}\right) ^{2}}{c_{Ai}^{2}}\right]
k_{r}^{2}c_{Ad}^{2},
\]

\noindent where $c_{Ad}=c\omega _{cd}/\omega _{pd}$ is the dust Alfv\'{e}n
velocity. Supposing that $|v_{i0}-v_{d0}|\gg (\nu _{nd}/\nu
_{in})^{1/2}c_{Ai}$ (the sign $|$ $|$ denotes an absolute value), the growth
rate of instability $\gamma $ will be equal to 
\begin{equation}
\gamma =\left( \frac{\nu _{in}}{\nu _{dn}}\right) ^{1/2}\frac{c_{Ad}}{c_{Ai}}%
k_{r}\left\vert v_{i0}-v_{d0}\right\vert .
\end{equation}

\noindent This solution exists if 
\[
\frac{\nu _{n}}{k_{r}}\left( \frac{\nu _{dn}}{\nu _{in}}\right) ^{1/2}\frac{%
c_{Ai}}{c_{Ad}}>\left\vert v_{i0}-v_{d0}\right\vert >c_{sn}\left( \frac{\nu
_{dn}}{\nu _{in}}\right) ^{1/2}\frac{c_{Ai}}{c_{Ad}}.
\]

\noindent From these inequalities, it follows that $\lambda _{r}\gg 2\pi
c_{sn}/\nu _{n}$, where $\lambda _{r}=2\pi /k_{r}$ is the wavelength of
perturbations. For parameters given above, we obtain $c_{sn}=1.73$ km s$^{-1}
$ $(\gamma _{a}=3)$, $c_{Ai}=600$ km s$^{-1}$, $c_{Ad}=6.8$ km s$^{-1}$, $%
\left\vert v_{i0}-v_{d0}\right\vert >10$ km s$^{-1}$, and $\lambda _{r}\gg
1.73\times 10^{10}$ km. The estimation of the growth rate (35) at $%
\left\vert v_{i0}-v_{d0}\right\vert =15$ km s$^{-1}$ and $\lambda
_{r}=5\times 10^{10}$ km is the following: $\gamma =3\times 10^{-10}$ s$^{-1}
$. The condition of magnetization (24) is satisfied for ions as well as for
dust grains. The case of unmagnetized dust grains has been considered in
(Nekrasov 2009 a)

In the case of thermal species and/or short wavelength perturbations, $\mu
_{n}k_{r}^{2}\gg \omega $ and $k_{r}^{2}c_{sj}^{2}\gg \omega \nu _{jn}$, we
consider the electron-ion plasma because conditions $k_{r}^{2}c_{sd}^{2}\gg
\omega \nu _{dn}$ and $k_{r}v_{d0}/\omega \ll \omega _{cd}/\nu _{dn}$ (see
inequalities (24)) are incompatible for solution (33) and parameters given
above. In the absence of dust grains, Equation (33) has the form (34). The
condition $k_{r}^{2}c_{si}^{2}\gg \gamma \nu _{in}$ and the right inequality
(24) for ions $k_{r}v_{i0}/\gamma \ll \omega _{ci}/\nu _{in}$ (for electrons
these conditions are weaker) are compatible if 
\[
\frac{m_{i}\nu _{in}}{m_{e}\nu _{en}}\ll \frac{\beta }{2\left( 1+\beta
\right) }\frac{\omega _{ci}^{2}}{\nu _{in}^{2}},
\]

\noindent where $\beta =12\pi n_{i}\left( T_{e}+T_{i}\right) /B_{0}^{2}$.
For number densities $n_{i}$ and $n_{n}$ used above, this inequality is also
not satisfied. At $\beta \ll 1$\ it can be written in the form,%
\[
\frac{\omega _{pi}}{\nu _{in}}\frac{c_{s}}{c}\gg 1.41\left( \frac{m_{i}\nu
_{in}}{m_{e}\nu _{en}}\right) ^{1/2}.
\]%
For $n_{n}\sim $\ $10^{4}$\ cm$^{-3}$\ and $n_{i}/n_{n}\sim 10^{-6}$,\ this
condition is satisfied. 

We have excluded drift waves in our study. This can be done if the frequency
spectra of these waves and perturbations under consideration are different.
As is known, the frequency of drift waves, for example, in the electron-ion
plasma, is equal to $\omega =\left( T_{e}/m_{i}\omega _{ci}n_{i}\right) %
\left[ \mathbf{k\times \nabla }n_{i}\right] _{z}$. For axisymmetric
perturbations, this frequency is equal to zero, if the number density is
uniform in the azimuthal direction. If $\partial n_{i}/n_{i}r\partial \theta
\sim L_{\theta }^{-1}\neq 0$\ in any region, then $\omega \sim $\ $\omega
_{ci}k_{r}\rho _{i}^{2}L_{\theta }^{-1}$. Comparing this expression, for
example, with the growth rate (34), we see that under condition 
\[
\left( \frac{\nu _{en}}{\omega _{ce}}\right) ^{1/2}\frac{|v_{e0}-v_{i0}|}{%
c_{Ai}}\gg \left( k_{r}L_{\theta }\right) ^{1/2}\frac{\rho _{i}}{L_{\theta }}
\]%
($\beta \ll 1$), the growth rate is larger than the drift frequency. The
last condition can easily be satisfied because $\rho _{i}$ is much smaller
than $L_{\theta }$.

\section{Conclusion}

In the present paper, we have studied electromagnetic streaming
instabilities\ in the thermal viscous regions of rotating astrophysical
objects, such as magnetized accretion disks, molecular clouds, their cores,
and elephant trunks. However, the results obtained can be applied to any
regions of interstellar medium, where different background velocities
between charged species can arise.

We have considered a weakly ionized multicomponent plasma consisting of
electrons, ions, dust grains, and neutrals. The cyclotron frequencies of
charged species have been supposed to be larger than their collisional
frequencies with neutrals. The axisymmetric perturbations across the
background magnetic field have been investigated. We have taken into account
the effect of perturbation of collisional frequencies due to density
perturbations of species. New compressible instabilities generated by the
different equilibrium velocities of species have been found in the cold and
thermal limits either when the viscosity of neutrals can be neglected or
when it is important. For the perturbations considered, the viscosity of
magnetized charged species is negligible.

In dense accretion disks, the ions are unmagnetized while the electrons
remain magnetized (e.g., Wardle \& Ng 1999). In this case, the viscosity of
ions can play the same role as that for neutrals. However, our present model
does not describe such objects.

The electromagnetic streaming instabilities studied in the present paper can
be a source of turbulence in weakly ionized magnetized astrophysical objects
in regions where thermal and viscous effects can play a role.

\bigskip 

The insightful and constructive comments and suggestions of the anonymous
referee are gratefully acknowledged.

\bigskip 

\paragraph{\noindent References\\}
\smallskip
\noindent \\
\noindent Abergel, A., Verstraete, L., Joblin, C., Laureijs, R., \&
Miville-Desch\^{e}nes, M.-A. 2005, Space Sci. Rev., 119, 247

\noindent Balbus, S. A., \& Henri, P. 2008, ApJ, 674, 408

\noindent Braginskii, S. I. 1965, Rev. Plasma Phys., 1, 205

\noindent Caselli, P., Walmsley, C. M., Terzieva, R., \& Herbst, E. 1998,
ApJ, 499, 234

\noindent Draine, B. T., Roberge, W. G., \& Dalgarno, A. 1983, ApJ, 264, 485

\noindent Masada, Y., \& Sano, T. 2008, ApJ, 689, 1234

\noindent Mendis, D. A., \& Rosenberg, M. 1994, Annu. Rev. Astron.
Astrophys., 32, 419

\noindent Nekrasov, A. K. 2007, Phys. Plasmas, 14, 062107

\noindent Nekrasov, A. K. 2008 a, Phys. Plasmas, 15, 032907

\noindent Nekrasov, A. K. 2008 b, Phys. Plasmas, 15, 102903

\noindent Nekrasov, A. K. 2009 a, Phys.Plasmas, 16, 032902

\noindent Nekrasov, A. K. 2009 b, ApJ, 695, 46

\noindent Nekrasov, A. K. 2009 c, ApJ, 704, 80

Pessah, M. E., \& Chan, C. 2008, ApJ, 684, 498

\noindent Ruffle, D. P., Hartquist, T. W., Rawlings, J. M. C., \& Williams,
D. A. 1998, A\&A, 334, 678

\noindent Wardle, M., \& Ng, C. 1999, MNRAS, 303, 239

\noindent Youdin, A. N., \& Goodman, J. 2005, ApJ, 620, 459

\end{document}